\documentstyle[12pt]{article}

\catcode`\@=11
\def\marginnote#1{}
\newcount\hour
\newcount\minute
\newtoks\amorpm
\hour=\time\divide\hour by60
\minute=\time{\multiply\hour by60 \global\advance\minute by-\hour}
\edef\standardtime{{\ifnum\hour<12 \global\amorpm={am}%
        \else\global\amorpm={pm}\advance\hour by-12 \fi
        \ifnum\hour=0 \hour=12 \fi
        \number\hour:\ifnum\minute<10 0\fi\number\minute\the\amorpm}}
\edef\militarytime{\number\hour:\ifnum\minute<10 0\fi\number\minute}
\def\draftlabel#1{{\@bsphack\if@filesw {\let\thepage\relax
   \xdef\@gtempa{\write\@auxout{\string
      \newlabel{#1}{{\@currentlabel}{\thepage}}}}}\@gtempa
   \if@nobreak \ifvmode\nobreak\fi\fi\fi\@esphack}
        \gdef\@eqnlabel{#1}}
\def\@eqnlabel{}
\def\@vacuum{}
\def\draftmarginnote#1{\marginpar{\raggedright\scriptsize\tt#1}}
\def\draft{\oddsidemargin -.5truein
        \def\@oddfoot{\sl preliminary draft \hfil
        \rm\thepage\hfil\sl\today\quad\militarytime}
        \let\@evenfoot\@oddfoot \overfullrule 3pt
        \let\label=\draftlabel
        \let\marginnote=\draftmarginnote
   \def\@eqnnum{(\theequation)\rlap{\kern\marginparsep\tt\@eqnlabel}%
\global\let\@eqnlabel\@vacuum}  }

\def\preprint{\twocolumn\sloppy\flushbottom\parindent 1em
        \leftmargini 2em\leftmarginv .5em\leftmarginvi .5em
        \oddsidemargin -.5in    \evensidemargin -.5in
        \columnsep 15mm \footheight 0pt
        \textwidth 250mmin      \topmargin  -.4in
        \headheight 12pt \topskip .4in
        \textheight 175mm
        \footskip 0pt
        \def\@oddhead{\thepage\hfil\addtocounter{page}{1}\thepage}
        \let\@evenhead\@oddhead \def\@oddfoot{} \def\@evenfoot{} }
\def\titlepage{\@restonecolfalse\if@twocolumn\@restonecoltrue\onecolumn
     \else \newpage \fi \thispagestyle{empty}\c@page\z@
        \def\thefootnote{\fnsymbol{footnote}} }

\def\endtitlepage{\if@restonecol\twocolumn \else  \fi
        \def\thefootnote{\arabic{footnote}}
        \setcounter{footnote}{0}}  

\catcode`@=12
\relax
\def\bea{\begin{array}}
\def\bem{\begin{displaymath}}
\def\beq{\begin{equation}}
\def\eea{\end{array}}
\def\eem{\end{displaymath}}
\def\eeq{\end{equation}}

\def\Im{\mathop{\rm Im}}
\def\NP#1#2#3{Nucl. Phys. \underline{#1} (19#2) #3}
\def\ov{\overline}

\def\PL#1#2#3{Phys. Lett. \underline{#1} (19#2) #3}
\def\PR#1#2#3{Phys. Rev. \underline{#1} (19#2) #3}

\def\Re{\mathop{\rm Re}}

\relax
\newcommand{\be}{\begin{equation}}
\newcommand{\en}{\end{equation}}
\newcommand{\ba}{\begin{eqnarray}}
\newcommand{\ea}{\end{eqnarray}}
\newcommand{\ee}{\end{equation}}

\def\crbig{\\\noalign{\vspace {3mm}}}

\def\slash #1{{\not \hspace{-.5mm}#1}}
\relax

\begin{document}
\topmargin-2.4cm
\renewcommand{\theequation}{\thesection.\arabic{equation}}
\begin{titlepage}
\begin{flushright}
{\small CERN--TH/99-254\\
LPTENS-99/28\\  
NEIP--99--015 \\
hep--th/9908137\\
August 1999}
\end{flushright}
\vspace{.4cm}
\begin{center}{\Large\bf
Non-perturbative Supersymmetry Breaking
and Finite Temperature Instabilities in N=4 Superstrings$^*$}

\vspace{.6cm}
{\bf I. Antoniadis$^{\,a}$,  J.-P. Derendinger$^{\, b}$
and C. Kounnas$^{\,c,\,d}$}
\vspace{.6cm}

{\small\sl
$^a$Centre de Physique Th{\'e}orique, Ecole Polytechnique $^\dagger$,
\\
F-91128 Palaiseau, France}\\ [3mm]
{\small\sl $^b$Institut de Physique, Universit\'e de
Neuch\^atel,\\
Breguet 1, CH-2000 Neuch\^atel, Switzerland}\\[3mm]
{\small\sl $^c$
Laboratoire de Physique Th\'eorique, Ecole Normale 
Sup\'erieure $^{\dagger\dagger}$, \\
24 rue Lhomond, F--75231 Paris Cedex 05, France}\\[3mm]
{\small\sl $^d$
Theory Division, CERN, 1211 Gen\`eve 23, Switzerland}
\end{center}
\vskip .3cm
\begin{center}
{\bf Abstract}
\end{center}
\begin{quote}
We obtain the non-perturbative effective potential for the dual
five-dimen\-sio\-nal $N=4$ strings in the context of finite-temperature
regarded as a breaking of supersymmetry into four space-time
dimensions. Using the properties of gauged $N=4$ supergravity we
derive the universal thermal effective potential describing all
possible high-temperature instabilities of the known $N=4$
superstrings.  These strings undergo a high-temperature transition to a
new phase in which five-branes condense. This phase is described in
detail, using both the effective supergravity and non-critical string
theory in six dimensions. In the new phase, supersymmetry is
perturbatively restored but broken at the non-perturbative level.
\end{quote}
\vspace{.5cm}
\begin{center}
{\it
To appear in the Proceedings of the Corfu Summer Institute on 
Elementary Particle Physics, Corfu, Greece, September 1998.
}
\end{center}
\vspace{.3cm}
\begin{flushleft}
\rule{8.1cm}{0.2mm}\\
$^{\star}$
{\small Research supported in part by
the EEC under the TMR contracts ERBFMRX-CT96-0090 and
ERBFMRX-CT96-0045, and by the Swiss National Science
Foundation and the Swiss Office for Education and Science.} \\
$^{\dagger}$
{\small Unit\'e mixte du CNRS UMR 7644.} \\
$^{\dagger\dagger}$ 
{\small Unit\'e mixte de Recherche UMR 8549, associ\'ee au CNRS
et \`a l'ENS.}
\end{flushleft}

\end{titlepage}
\setcounter{footnote}{0}
\setcounter{page}{0}
\setlength{\baselineskip}{.7cm}
\newpage
%
%

\section{Introduction}
A convenient way to analyse a $D$-dimensional theory at
finite temperature is to identify 
the temperature with the inverse
radius of a compactified Euclidean time on $S^1$, $R=1/2\pi T$
and to modify the boundary conditions around the $S^1$
according to spin-statistics: periodic for bosons, 
antiperiodic for fermions. The modified boundary conditions
shift the $S^1$ Kaluza-Klein charge by an amount proportional to
the helicity of the state, $m\rightarrow m+Q$. In string theories
this  shift is generalised and includes a winding contribution:
$m\rightarrow m+Q+\delta n/2$. This shift is dictated by the
world-sheet
modular invariance; $\delta=1$ for the heterotic string and  $\delta=0$
for the type II strings \cite{AW, KR, AK}. Furthermore, the GSO
projection in the odd winding number sector is reversed.

For an even winding number $n$, the
thermal modification  can be
regarded as a shift of $m$ and $Q$ compatible with the
(supersymmetric) GSO projection. As a consequence, the spectrum in
even $n$ sectors is not different in the thermal and supersymmetric
cases, the mass formula for the (lightest) BPS fermions, gauge bosons
and scalars with even windings $n$ remains ${\cal M}^2 = P^2$, with
$m$ modified, and tachyonic states are not
present. The situation is not the same for states with odd winding
number $n$ due to the reversion of the GSO projection. It follows that
the only states that can become tachyonic are those with $n=\pm 1$ and
correspond to $(D-1)$-dimensional scalars coming 
from the longitudinal components of the
$D$-dimensional metric.

Tachyons cannot appear in a perturbative
supersymmetric field theory, which behaves like the zero-winding
sector of strings; all (squa\-red) masses are increased by finite
temperature corrections, ${\cal M}^2 = P^2$, and a thermal
instability is never generated by a state becoming tachyonic at high
temperature. However, as we will see below, in
non-perturbative
supersymmetric field theories such an instability can arise from
thermal dyonic modes, which behave 
as the odd winding string states \cite{ADK}.
Indeed, in theories with $N=4$ supersymmetries, the BPS mass formula
is determined by the central extension of the corresponding
superalgebra \cite{N=4BPS}--\cite{KK} and dyonic field theory states
are
mapped to string winding modes \cite{Sen, KK}. Using heterotic--type
II duality, one can argue that the thermal shift of the BPS masses
modifies only the perturbative momentum charge $m$. In both heterotic
and type II perturbative strings, the thermal winding number $n$ is
not affected by the temperature shifts.
Since, in dimensions lower than six, heterotic--type II duality
exchanges the winding numbers $n$ of the two theories, and since the
winding number of the one theory is the magnetic charge of the other,
it is inferred that field theory magnetic numbers are not shifted at
finite temperature. This in turn indicates how to modify the BPS mass
formula at finite temperature \cite{ADK}.

It turns out that string theories with $D$-di\-men\-sional space-time
supersymmetry look at finite temperature as if supersymmetry were
spontaneously broken in $D-1$ dimensions \cite{AW}--\cite{ADK}.

\section{Thermal masses and string-string dualities}

The non-perturbative four-dimensional thermal mass formula has been
obtained in Ref. \cite{ADK}. The procedure is to start with the $N=4$
four-dimen\-sio\-nal BPS mass formula on a circle with radius $R$,
which depends on an effective string tension
\beq
\begin{array}{rcl}
\label{Tpqris}
T_{p,q,r} &=& \displaystyle{{p\over\alpha_H^\prime}
+{q\over\lambda_H^2\alpha_H^\prime}
+{r R_6^2\over\lambda_H^2(\alpha_H^\prime)^2}} \crbig
&=& \displaystyle{
{p\over\alpha_H^\prime}
+{q\over\alpha_{IIA}^\prime}
+{r\over\alpha_{IIB}^\prime}.}
\end{array}
\eeq
The modified finite-temperature formula reads then
\beq
\label{mass5}
{\cal M}^2_T = \displaystyle{
\left({m+Q'+{kp\over 2}\over R}+
k~T_{p,q,r}~R\right)^2} 
\displaystyle{ -2 ~T_{p,q,r}~\delta_{|k|,1}
{}~\delta_{Q',0}\, ,}
\eeq
In these expressions, $kp$ is the winding number in the heterotic
string representation with intercept scale $\alpha^\prime_H$,
while $kq$ is the magnetic Kaluza-Klein charge and $kr$ is the
magnetic winding charge. Still in the heterotic picture, $kq$ is
the wrapping number of
the heterotic five-brane around $T^4\times S^1_R$,
while $kr$ corresponds to the same
wrapping number after performing a T-duality along the circle
of the sixth dimension.
The shift in the momentum Kaluza-Klein number $m$
$$
m  \quad\longrightarrow\quad m+Q^\prime + {kp\over2},
$$
is dictated by the change of boundary conditions at finite temperature,
compared to a simple circle compactification. The helicity charge
$Q^\prime$
distinguishes (four-dimensional) bosons and fermions. The shift $kp/2$
is dictated then by modular invariance of the dual perturbative
strings.
Finally, the mass formula (\ref{mass5}) includes a subtraction
in the odd $k$ winding sector of the effective $T_{p,q,r}$ string.
We refer to Ref. \cite{ADK} for a detailed discussion.

The mass formula (\ref{mass5}) depends on three parameters: the
six-dimensional heterotic string coupling $\lambda_H$, the circle
compactification $R_6$ from six to five dimensions, and the radius $R$
which will be identified with the inverse temperature. It also depends
on
a scale: the duality invariant scale is the (four-dimensional)
Planck scale $\kappa=\sqrt{8\pi}M_P^{-1}=
(2.4\times10^{18}\,\,{\rm GeV})^{-1}$. It is convenient to introduce
instead
of $\lambda_H$, $R_6$ and $R$ the (dimensionless) variables
\beq
\label{stuare}
t = {RR_6\over\alpha^\prime_H},
\qquad u = {R\over R_6}, \qquad  s= g_H^{-2} = {t\over\lambda_H^2},
\eeq
which will be directly related to moduli of the effective supergravity
theory; $g_H$ is now the {\it four}-dimensional heterotic string
coupling. The various $\alpha^\prime$ scales in the effective 
tension
(\ref{Tpqris}) are
\beq
\label{alphaare}
\alpha_H^\prime = 2\kappa^2 s, \quad
\alpha_{IIA}^\prime = 2\kappa^2 t, \quad
\alpha_{IIB}^\prime = 2\kappa^2 u, \quad
\eeq
when expressed in Planck units. In addition, string--string dualities
have a simple formulation using these variables.
Before the temperature shift on $m$, the BPS mass formula
is invariant under  the exchanges
$s\leftrightarrow t$, $s\leftrightarrow u$ and $t\leftrightarrow u$.
These operations correspond respectively to heterotic--IIA, 
heterotic--IIB and IIA--IIB dualities in the undeformed (by temperature)
$N=4$ supersymmetric theories.
In terms of $s$, $t$ and $u$, the
temperature radius $R$ is given by
\beq
\label{Ris}
R^2 = \alpha^\prime_H tu  = 2\kappa^2 stu
\eeq
and $R$ is by construction identical in all three string theories.

As a consequence of the BPS conditions and
the $s\leftrightarrow t\leftrightarrow u$ duality symmetry in the
undeformed supersymmetric theory,
the integers $p$, $q$, $r$ are non-negative and relatively prime.
Furthermore, $mk\ge -1$ because of
the inversion of the GSO projection in the theory deformed by
temperature. Using these constraints, it is straightforward to show
that
in general there are two potential tachyonic series with $m=-1$ and
$p=1, 2$:
\beq
\label{Tpqr2}
\begin{array}{rl}
p=1, \quad \forall  (q,r) ~ {\rm relat. ~primes}:&\qquad
R = \displaystyle{\left({\sqrt2\pm1 \over
\sqrt2}\right) {1\over\sqrt{T_{1,q,r}}}}, \crbig
p=2, \quad \forall  (p,q,r) ~{\rm relat. ~primes}:&\qquad
R = \displaystyle{\sqrt{2\over T_{2,q,r}}}
\end{array}
\eeq
One of the perturbative heterotic, type IIA or type IIB potential
tachyons corresponds to a critical temperature that is always lower
than the above two series. The perturbative Hagedorn temperatures
are:
$$
\begin{array}{c}
{\rm heterotic \,\,  tachyon:}\quad
m=\mp1, kp =\pm1, Q^\prime =0
\crbig
2\pi T = \left({\sqrt2-1}\right)
\displaystyle{\sqrt{2\over\alpha^\prime_H}};
\crbig
{\rm type \,\, IIA \,\, tachyon:}\quad
m=0, kq=\pm1, Q^\prime=0 \crbig
2\pi T = \displaystyle{1\over\sqrt{2\alpha^\prime_{IIA}}};
\crbig
{\rm type \,\, IIB \,\, tachyon:}\quad
m=0, kr=\pm1, Q^\prime=0 \crbig
2\pi T = \displaystyle{1\over\sqrt{2\alpha^\prime_{IIB}}},
\end{array}
$$
and $T= (2\pi R)^{-1}$.

This discussion shows that the temperature modification of the mass
formula inferred from perturbative strings and applied to the
non-per\-tur\-ba\-tive BPS mass formula produces the appropriate
instabilities in terms of Hagedorn temperature. We will now proceed
to show that it is possible to go beyond the simple enumeration of
Hagedorn temperatures. We will construct an effective supergravity
Lagrangian that allows a study of the nature of the non-perturbative
instabilities and the dynamics of the various thermal phases.

The above formula hold for supersymmetry broken by temperature effects
in Euclidean spa\-ce. They would similarly hold for a
non-super\-sym\-metric
four-dimensional Minkowski theory in which supersymmetry 
would be broken
by a particular Scherk-Schwarz compactification of the fifth dimension.

\section{Effective supergravity in  N=1 representation}

In the previous section, we have studied the appearance of tachyonic 
states generating thermal instabilities
at the level of the mass
formula for $N=4$ BPS states. To obtain information on dynamical
aspects of these instabilities, we now construct the full
temperature-dependent effective potential for the
would-be tachyonic states.

Our procedure to construct the effective theory is as follows. We
consider five-dimensional $N=4$ theories at finite temperature.
They can then effectively be described by four-dimensional theories,
in which supersymmetry is spontane\-ous\-ly broken by thermal effects.
Since we want to limit ourselves to the description of instabilities,
it is sufficient to only retain, in the full $N=4$ spectrum, the
potentially massless and tachyonic states. This restriction will lead
us to consider only spin 0 and 1/2 states, the graviton and the
gravitino\footnote{The four gravitinos remain degenerate at
finite temperature; it is then sufficient to retain only one of
them.}. This sub-spectrum is described by an $N=1$ supergravity
with chiral multiplets.

The scalar manifold of a generic, {\it unbroken}, $N=4$ theory is
\cite{DF}--\cite{FK}
\beq
\label{manif1}
\begin{array}{l}
\displaystyle\left({Sl(2,R) \over U(1)}\right)_S \times \, G/H, \crbig
G/H = \displaystyle
\left({SO(6,r+n)\over SO(6)\times SO(r+n)}\right)_{T_I,\phi_A}.
\end{array}
\eeq
The manifold $G/H$ of the $N=4$ vector multiplets
naturally splits into a part that includes the $6r$
moduli $T_I$, and a second part which includes the infinite number
$n\rightarrow\infty$ of BPS states $\phi_A$.

In the manifold $G/H$, we are only interested in keeping the six BPS
states $Z_A^\pm$, $A=1,2,3$, which, according to our discussion in
the previous section, generate thermal instabilities in heterotic,
IIA and IIB strings. For consistency, these states must be
supplemented by two moduli $T$ and $U$ among the $T_I$'s. We consider
heterotic and type II strings respectively on $T^4\times S^1_6\times
S^1_5$ and $K_3\times S^1_6\times S^1_5$, where $S^1_6$ is a trivial
circle and $S_5^1$ is the temperature circle. The moduli $T$ and $U$
describe the $T^2\equiv S^1_5\times S^2_6$ torus. Thus, $r+n = 8$ in
the $N=4$ manifold (\ref{manif1}). To construct the appropriate
truncation of the scalar manifold $G/H$, which only retains the
desired states of $N=1$ chiral multiplets, we use a $Z_2\times Z_2$
subgroup contained in the $SO(6)$ R-symmetry of the coset $G/H$. This
symmetry can be used as the point group of an $N=1$ orbifold
compactification, but we will only use it for projecting out
non-invariant states of the $N=4$ theory\footnote{Only
untwisted states would contribute to thermal instabilities.} with
$r+n=8$.

The $Z_2\times Z_2$ projection splits $H=SO(6)\times SO(8)$ in
$SO(2)^3\times SO(2)\times SO(3)^2$ and the scalar manifold
becomes
\beq
\label{manif3}
\begin{array}{l}
\left({Sl(2,R) \over U(1)}\right)_S \times
\left({Sl(2,R) \over U(1)}\right)_T \times
\left({Sl(2,R) \over U(1)}\right)_U \crbig
\hspace{.6cm}
\times
\left({SO(2,3)\over SO(2)\times SO(3)}\right)_{Z_A^+}\times
\left({SO(2,3)\over SO(2)\times SO(3)}\right)_{Z_A^-},
\end{array}
\eeq
$A=1,2,3$. The tachyonic instabilities will however be controlled by
the diagonal sub-manifold,
\beq
\label{manif}
\begin{array}{l}
\displaystyle{\left({Sl(2,R) \over U(1)}\right)_S \times
\left({Sl(2,R) \over U(1)}\right)_T \times
\left({Sl(2,R) \over U(1)}\right)_U }\crbig
\hspace{.5cm}
\displaystyle{\times\left({SO(2,3)\over SO(2)\times
SO(3)}\right)_{Z_A},}
\end{array}
\eeq
identifying $Z_A^+=Z_A^-=Z_A$.

From the structure of the truncated scalar manifold, we find
that the K\"ahler potential is
\beq
\label{Kis}
\begin{array}{rcl}
K &=& -\log[(S+S^*)(T+T^*)(U+U^*)] \crbig
&&-2\log [1 -2Z_AZ_A^{*} + (Z_A Z_A)(Z_B^{*}Z_B^{*})].
\end{array}
\eeq
The superpotential of the theory is obtained using the fact that at the
level of $N=4$ supergravity, finite
temperature corresponds to a particular Scherk-Schwarz gauging,
breaking supersymmetry spontaneously. This gauging is defined by a
set of generalized (field-dependent) structure constants,
involving the compensating multiplets which are used to define
the $G/H$ manifold. The truncation to $N=1$
supergravity delivers then the following expression for the
superpotential:
\beq
\label{Wis}
W = \sqrt2 \bigl[ (1-Z_AZ_A)(1-Z_BZ_B)
 +2(TU-1)Z_1^2 + 2SUZ_2^2 +
2STZ_3^2  \bigr].
\eeq

From the K\"ahler potential and the superpotential we can then compute
the full effective scalar potential and study its instabilities. Its
complicated expression simplifies drastically 
in the directions relevant to instabilities.
Introducing the variables
\beq
\label{variables}
\begin{array}{c}
s=\Re S, \qquad t=\Re T, \qquad u =\Re U, \crbig
z_A= \Re Z_A , \quad  x^2 = \sum_A z_A^2, \qquad
H_A = \displaystyle{z_A\over 1-x^2}, \qquad A=1,2,3, \crbig
\xi_1 = tu, \qquad \xi_2 = su, \qquad \xi_3 = st ,
\end{array}
\eeq
the resulting scalar potential becomes
\beq
\label{pot4}
\begin{array}{rcl}
V &=& V_1 + V_2 + V_3, \crbig
\kappa^4 V_1 &=& \displaystyle{4\over s}\Bigl[
(\xi_1+\xi_1^{-1})H_1^4  
\displaystyle{
+{1\over4}(\xi_1-6+\xi_1^{-1})H_1^2 \Bigr], }
\crbig
\kappa^4 V_2 &=& \displaystyle{4\over t}
\Bigl[ \xi_2H_2^4 +{1\over4}(\xi_2-4)H_2^2\Bigr],
\crbig
\kappa^4 V_3 &=& \displaystyle{4\over u}
\Bigl[ \xi_3H_3^4 +{1\over4}(\xi_3-4)H_3^2\Bigr].
\end{array}
\eeq
This expression displays the duality properties
\begin{itemize}
\item[] $\xi_1 \,\,\rightarrow\,\, \xi_1^{-1}$:
heterotic temperature duality;
\item[] $t \,\,\leftrightarrow\,\, u$, $H_2 \,\,\leftrightarrow\,\,
H_3$:
IIA--IIB duality.
\end{itemize}

\subsection{Phase structure of the thermal effective
theory}\label{secphase}

The scalar potential (\ref{pot4}) derived from our effective
supergravity possesses four different phases corresponding to
specific regions of the $s$, $t$ and $u$ moduli space. Their
boundaries are defined by critical values of the moduli $s$, $t$, and
$u$ (or of $\xi_i$, $i=1,2,3$), or equivalently by critical values of
the temperature, the (four-dimensional) string coupling and the
compactification radius $R_6$. These four phases are:
\begin{enumerate}
\item The {\sl low-temperature} phase: \\
$T<(\sqrt2-1)^{1/2}/(4\pi\kappa)$;
\item The {\sl high-temperature heterotic} phase: \\
$T>(\sqrt2-1)^{1/2}/(4\pi\kappa)$, $\,\,g_H^2<(2+\sqrt2)/4$;
\item The {\sl high-temperature type IIA} phase: \\
$T>(\sqrt2-1)^{1/2}/(4\pi\kappa)$, $\,\,g_H^2>(2+\sqrt2)/4\,\,$
and $\,\,R_6>\sqrt{\alpha^\prime_H}$;
\item The {\sl high-temperature type IIB} phase: \\
$T>(\sqrt2-1)^{1/2}/(4\pi\kappa)$, $\,\,g_H^2>(2+\sqrt2)/4\,\,$
and $\,\,R_6<\sqrt{\alpha^\prime_H}$.
\end{enumerate}
The distinction between phases 3 and 4 is, however, somewhat
academic,
since there is no phase boundary at $R_6=\sqrt{\alpha^\prime_H}$.

\subsubsection{Low-temperature phase}

This phase, which is {\it common} to all three strings,
is characterized by
\beq
\label{ltp1}
H_1=H_2=H_3=0, \qquad V_1=V_2=V_3=0.
\eeq 
The potential vanishes for all values of the moduli $s$, $t$ and
$u$, which are then restricted only by the stability of the phase,
namely the absence of tachyons in the mass spectrum of the scalars
$H_i$. This mass spectrum is analysed in Ref. \cite{ADK}. Stability 
requires then:
\beq
\label{ltp2}
\xi_1 > \xi_H = (\sqrt2+1)^2, \qquad
\xi_2 >  4, \qquad
\xi_3 >  4.
\eeq
From the above conditions, it follows in particular that the
temperature must verify
\beq
\label{Tis}
T = {1\over 2\pi\kappa} \left({1\over\xi_1\xi_2\xi_3}\right)^{1/4}
< {(\sqrt2-1)^{1/2}\over4\pi\kappa}.
\eeq
Since the (four-dimensional) string couplings are
$$
s = \sqrt2 g_H^{-2}, \qquad\quad t = \sqrt 2 g_A^{-2}, \qquad\quad
u = \sqrt2 g_B^{-2},
$$
this phase exists in the perturbative regime of all three strings.
The relevant light thermal states are just the massless modes of the
five-dimen\-sio\-nal $N=4$ supergravity, with thermal mass scaling like
$1/R \sim T$.

Alternatively, if this effective theory is considered as a
six-dimensional
Minkowski model compactified on $S^1_R\times S^1_{R_6}$, with
spontaneously broken supersymmetry, then the
lowest $S^1_R$ Kalu\-za-Klein modes have masses shifted by a
quantity proportional to the gravitino mass scale,
$$
m^2_{3/2} = {1\over 4R^2},
$$
which then controls both the Kaluza-Klein mass shifts and the splitting
of supersymmetric multiplets.

\subsubsection{High-temperature heterotic phase}

This phase is defined by
\beq
\label{hhp1}
\xi_H > \xi_1 > {1\over\xi_H},
\quad \xi_2>4, \quad \xi_3>4,
\eeq
with $\xi_H =(\sqrt2+1)^2$, as in Eq. (\ref{ltp2}). The inequalities
on $\xi_2$ and $\xi_3$ eliminate type II instabilities. In this
region of the moduli, and after minimization with respect to $H_1$,
$H_2$ and $H_3$, the potential becomes
$$
\kappa^4 V = -{1\over s}{(\xi_1+\xi_1^{-1}-6)^2\over
16(\xi_1+\xi_1^{-1})}.
$$
It has a stable minimum for fixed $s$ (for fixed $\alpha^\prime_H$)
at the minimum of the self-dual\footnote{With respect to heterotic
temperature duality.} quantity $\xi_1+\xi_1^{-1}$:
\beq
\label{highhetmin}
\begin{array}{c}
\xi_1=1 , \quad H_1={1\over2} , \quad H_2=H_3=0, \crbig
\kappa^4 V= -\displaystyle{1\over 2s}.
\end{array}
\eeq
The transition from the low-temperature vacuum is due to a
condensation of the heterotic thermal winding mode $H_1$, or
equivalently by a condensation of type IIA NS five-brane in the type
IIA picture.

At the level of the potential only, this phase exhibits a runaway
behaviour in $s$. We will show in the next section the existence
of a stable solution to the effective action  
with non-trivial metric and/or dilaton.

In heterotic language, $s$, $t$ and $u$ are particular combinations
of the four-dimensional gauge coupling $g_H$, the temperature $T=
(2\pi R)^{-1}$ and the compactification radius from six to five
dimensions $R_6$. The relations are
\beq
\label{hetstuare}
\begin{array}{c} \displaystyle{
s= \sqrt2g_H^{-2}, \quad t = \sqrt2{RR_6\over\alpha^\prime_H},
\quad u = \sqrt2{R\over R_6},}
\crbig \displaystyle{
\xi_1 = tu = {2R^2\over\alpha^\prime_H}, \quad
\xi_2 = {2R\over g_H^2R_6}, \quad
\xi_3 = {2RR_6\over\alpha^\prime_H g_H^2}. }
\end{array}
\eeq
As expected, $\xi_2$ and $\xi_3$ are related by radius inversion,
$R_6 \,\rightarrow\, \alpha^\prime_H R_6^{-1}$.
Then, in Planck units,
\beq
\label{hetTRuare}
\begin{array}{rcl}
R &=& \displaystyle{{1\over 2\pi T} = \kappa \sqrt{stu} =
\kappa[\xi_1\xi_2\xi_3]^{1/4},}
\crbig
R_6 &=& \displaystyle{
\kappa \left({2st\over u}\right)^{1/2}
= {\sqrt2 \kappa\xi_3\over[\xi_1\xi_2\xi_3]^{1/4}}.}
\end{array}
\eeq
The first equation indicates that the temperature, when expressed in
units of the {\it four-dimen\-sio\-nal} gravitational coupling constant
$\kappa$ is invariant under string--string dualities.

In terms of heterotic variables, the critical temperatures
(\ref{hhp1}) separating the heterotic pha\-ses are
\beq
\label{Thetcrit}
\begin{array}{rcrrcl}
\xi_1 &=& \xi_H: \quad&
2\pi T^<_H &=& \displaystyle{g_H\over 2^{1/4}\kappa}(\sqrt2-1),
\crbig
\xi_1 &=& \displaystyle{1\over\xi_H}:\quad &
2\pi T^>_H &=& \displaystyle{g_H\over 2^{1/4}\kappa}(\sqrt2+1).
\end{array}
\eeq
In addition, heterotic phases are separated from type II
instabilities by the following critical temperatures:
\beq
\label{typeIIcrit}
\begin{array}{rrclrcl}
{\rm IIA:}\quad& \xi_2&=&4,\quad&2\pi T_{A} &=&
\displaystyle{R_6\over 4\sqrt2\kappa^2},
\crbig
{\rm IIB:}\quad& \xi_3&=&4,\quad&2\pi T_{B} &=&
\displaystyle{1\over 2 g_H^2 R_6}.
\end{array}
\eeq
Then the domain of the moduli space that avoids type II
instabilities is defined by the inequalities $\xi_{2,3}>4$.
In heterotic variables,
\beq
\label{Tineq}
\begin{array}{rl}
2\pi T <& \displaystyle{
{1\over 2\alpha^\prime_H g_H^2}\, {\rm min}\left(
R_6\,\,;\,\,
\alpha^\prime_H/R_6\right) }\crbig
&= \displaystyle{{1\over 4\sqrt2\kappa^2}\, {\rm min}\left( R_6\,\,;\,\,
\alpha^\prime_H/R_6\right).}
\end{array}
\eeq
Type II instabilities are unavoidable when $T>T_{\rm self-dual}$,
with
$$
2\pi T_{\rm self-dual} = {1\over 2g_H^2\sqrt{\alpha^\prime_H}}
= {2^{1/4}\over 4\kappa g_H}.
$$
The high-temperature heterotic phase cannot be reached\footnote{From
low heterotic temperature.} for any value of the radius $R_6$ if
$$
T^<_H > T_{\rm self-dual},
$$
or
\beq
\label{glim}
g_H^2 > {\sqrt2+1\over2\sqrt2} \,\sim \,0.8536.
\eeq
In this case, $T^<_H$ always exceeds $T_{A}$ and $T_{B}$. Only type
II thermal instabilities exist in this strong-coupling regime and the
value of $R_6/\sqrt{\alpha^\prime_H}$ decides whether the type IIA or
IIB instability will have the lowest critical temperature, following
Eq. (\ref{typeIIcrit}).

If on the other hand the heterotic string is weakly coupled,
\beq
\label{glim1}
g_H^2 < {\sqrt2+1\over2\sqrt2},
\eeq
the high-temperature heterotic phase is reached for values of the
radius $R_6$ verifying $T^<_H < T_{A}$ and $T^<_H < T_{B}$, or
\beq
\label{R6ineq}
2\sqrt2g_H^2(\sqrt2-1) < {R_6\over\sqrt{\alpha^\prime_H}} <
{1\over2\sqrt2g_H^2(\sqrt2-1)}.
\eeq
The large and small $R_6$ limits, with fixed coupling $g_H$, again
lead to either type IIA or type IIB instability.

\subsubsection{High-temperature type IIA and IIB phases}

These phases are defined by inequalities:
\beq
\label{IIphases1}
\xi_2<4 \qquad {\rm and/or}\qquad \xi_3<4.
\eeq
In this region of the parameter space, either
$H_2$ or $H_3$ become tachyonic and acquire a vacuum value:
\beq
\label{IIAphase}
H_2^2 = {4-\xi_2\over 8\xi_2}, \qquad\qquad
\kappa^4 V_2=-{1\over t}{(4-\xi_2)^2\over16\xi_2},
\eeq
and/or
\beq
\label{IIBphase}
H_3^2 = {4-\xi_3\over 8\xi_3}, \qquad\qquad
\kappa^4 V_3=-{1\over u}{(4-\xi_3)^2\over16\xi_3}.
\eeq
In contrast with the high-temperature heterotic phase, the potential
does not possess stationary values of $\xi_2$ and/or $\xi_3$, besides
the critical $\xi_{2,3}=4$.

Suppose for instance that $\xi_2<4$ and $\xi_3>4$. The resulting
potential is then $V_2$ only and $\xi_2$ slides to zero. In this
limit,
$$
V= -\,{1\over stu\kappa^4},
$$
and the dynamics of $\phi \equiv -\log(stu)$ is described by the
effective Lagrangian
$$
{\cal L}_{\rm eff} = -{e\over2\kappa^2}\left[R
+{1\over6}(\partial_\mu\phi)^2 - {2\over\kappa^2}e^{\phi}\right].
$$
Other scalar components $\log(t/u)$ and $\log(s/u)$ have only
derivative couplings, since the potential only depends on $\phi$.
They can be taken to be constant and arbitrary. The dynamics only
restricts the temperature radius $\kappa^{-2}R^2=e^{-\phi}$, $R_6$
and the string coupling are not constrained, besides inequalities
(\ref{IIphases1}).

In conformally flat gravity background, the equation of motion of
the scalar $\phi$ is
$$
\hat{\Box}\phi =-{6\over\kappa^2}e^\phi.
$$
The solution of the above and the Einstein equations defines a
non-trivial gravitational $\phi$-back\-ground. This solution will
correspond to the high-temperature type II vacuum. We will not study
this solution further here.

\section{High-temperature heterotic phase}

The thermal phase relevant to weakly-coupled, high-temperature
heterotic strings at intermediate values of the radius $R_6$ [see
inequalities (\ref{glim1}) and (\ref{R6ineq})] has an interesting
interpretation; we study this here, using the information contained
in its effective theory, which is characterized by
Eqs. (\ref{highhetmin}):
\beq
\label{hethighvac}
tu = 1, \qquad H_1 = {1\over2}, \qquad H_2 = H_3 = 0.
\eeq
These values solve the equations of motion of all scalar
fields with the exception of $s$.
The resulting bosonic effective Lagrangian describing the dynamics of
$s$ and $g_{\mu\nu}$ is
\beq
\label{hethigh1}
{\cal L}_{\rm bos} = -{1\over2\kappa^2}eR - {e\over4\kappa^2}
(\partial_\mu\ln s)^2 + {e\over2\kappa^4s}.
\eeq
For all (fixed) values of $s$, the cosmological constant is negative
since $e^{-1}V=-(2\kappa^4s)^{-1}$ and the apparent
geometry is anti-de Sitter.
But the effective theory (\ref{hethighvac}) does not stabilize $s$.

To study the bosonic Lagrangian, we first rewrite it in the string
frame. Defining the dilaton as
\beq
\label{dilis}
e^{-2\phi} = s,
\eeq
and rescaling the metric according to
\beq
\label{grescal}
g_{\mu\nu} \quad\longrightarrow\quad
{2\kappa^2\over\alpha^\prime_H}e^{-2\phi} g_{\mu\nu},
\eeq
one obtains\footnote{Since the rescaling $g_{\mu\nu}\rightarrow
e^{-2\sigma}g_{\mu\nu}$ leads to
$e[R+6(\partial_\mu\sigma)^2]\rightarrow
e^{-2\sigma}eR$.}
\beq
\label{stframe}
{\cal L}_{\rm string~frame} =
{e^{-2\phi}\over\alpha^\prime_H}
\left[-eR+4e(\partial_\mu\phi)(\partial^\mu\phi)  
+{2e \over \alpha^\prime_H} \right].
\eeq
The equation of motion for the dilaton then is
\beq
\label{dileom}
R +4(\partial_\mu\phi)(\partial^\mu\phi) -4\Box\phi =
{2\over\alpha^\prime_H}.
\eeq
Comparing with the two-dimensional sigma-mo\-del dilaton
$\beta$-function \cite{beta} with central charge deficit
$\delta c = D-26$, which leads to
\beq
\label{dileom2}
R +4(\partial_\mu\phi)(\partial^\mu\phi) -4\Box\phi =
- {\delta c\over3\alpha^\prime_H},
\eeq
we find a central charge deficit $\delta c = -6$, or, for a
superstring,
\beq
\label{deltac}
\delta\hat c = {2\over3}\delta c = -4.
\eeq
In the string frame, a background for theory (\ref{stframe}) has flat
(sigma-model) metric $g_{\mu\nu}=\eta_{\mu\nu}$ and
linear dilaton dependence \cite{ABEN} on a spatial coordinate, say
$x^1$:
\beq
\label{lindil}
\phi = \phi_0 + Q x^1, \qquad\qquad
Q^2 = {\delta\hat c\over 8\alpha^\prime_H}={1\over2\alpha^\prime_H}
\eeq
($\phi_0$ is a constant).

The linear dilaton background breaks both four-dimensional Lorentz
symmetry and four-di\-men\-sio\-nal Poincar\'e supersymmetry. Since
supersymmetry breaks sponta\-neous\-ly, one expects to find goldstino
states in the fermionic mass spectrum and massive spin 3/2 states.
And, because of the non-trivial background, the theory in the
high-temperature heterotic phase is effectively a three-dimensional
supergravity.

To discuss the pattern of goldstino states, observe first that
the supergravity extension of the bosonic Lagrangian
(\ref{hethigh1}) includes a non-zero gravitino mass
term for all values of $s$ since
\beq
\label{m3/2}
m_{3/2}^2 = \kappa^{-2}\, e^{\cal G} = {1\over4\kappa^2 s} =
{1\over2\alpha^\prime_H} = Q^2.
\eeq
Notice also that the potential at the vacuum verifies
\beq
\label{potvac}
V = -{2\over\kappa^4}e^{\cal G} = -{1\over2\kappa^4 s}
= -{2\over \kappa^2}\,m_{3/2}^2.
\eeq
Consider then the transformation of fer\-mions in the chiral multiplet
$(z^i,\chi^i)$ \footnote{The notation is as in Ref. \cite{CFGVP},
with sign-reversed ${\cal G}$ and $\sigma^{\mu\nu} =
{1\over4}[\gamma^\mu,\gamma^\nu]$. Indices $i,j,\ldots$, enumerate
all chiral multiplets $(z^i,\chi^i)$.}:
\beq
\label{susytransf}
\delta\chi_{Li} = {1\over2}\kappa (\slash\partial z_i)\epsilon_R
-{1\over2}e^{{\cal G}/2}\,({\cal G}^{-1})_i^j{\cal G}_j \,\epsilon_L
+\ldots,
\eeq
omitting fermion contributions. In the high-tem\-pe\-ra\-ture
heterotic phase,
\beq
\label{Gmin}
{\cal G}_S= {\partial\over\partial S}{\cal G} = -{1\over2s},
\qquad\qquad
{\cal G}_a = {\partial\over\partial z^a}{\cal G} =0,
\eeq
and the K\"ahler metric is diagonal with
${\cal G}^S_S = (2s)^{-2}$. Since also
$$
\slash\partial s = -2Q s\gamma^1, \qquad e^{{\cal G}/2} = \kappa Q,
$$
only the fermionic partner $\chi_s$ of the dilaton $s$
participates in supersymmetry breaking, with the transformation
\beq
\label{susytranf2}
\delta \chi_s = {\sqrt s\over2}(1-\gamma^1)\epsilon.
\eeq
Supersymmetries generated by $(1-\gamma^1)\epsilon$ are then broken
in the linear dilaton background in the $x_1$ direction while those
with parameters $(1+\gamma^1)\epsilon$ remain unbroken. Starting then
from sixteen supercharges ($N=4$ supersymmetry) at zero
temperature, the high-temperature heterotic vacuum has eight unbroken
supercharges. Since the effective space-time symmetry is
three-dimensional, the high-temperature phase has $N_3=4$
supersymmetry: the linear dilaton background acts identically with
respect to the $N=4$ spinorial charges. It simply breaks one half
of the charges in each spinor. Thus, the high-temperature pha\-se is
expected to be stable because of supersymmetry of its effective field
theory and because of its superconformal content.

The mass spectrum of the effective supergravity theory in the linear
dilaton background is analyzed in Ref. \cite{ADK}. One first observes
that the K\"ahler potential does not induce any mixing between the
dilaton
multiplet and other chiral multiplets. Then, the dilaton multiplet only
plays an active role in the breaking of supersymmetry.

This splitting of chiral
multiplets does not exist in the
low-temperature phase $H_1=H_2=H_3=0$, in which
\beq
\label{lowphase1}
{\cal G}_S=-(2s)^{-1},\qquad
{\cal G}_T=-(2t)^{-1},\qquad
{\cal G}_U=-(2u)^{-1},
\eeq
with
$$
\psi_G = {1\over2s}\chi_s + {1\over2t}\chi_t + {1\over2u}\chi_u
$$
as goldstino state\footnote{Expressed using non-normalized fermions.
Canonical normalization of the spinors would lead to
$\psi_G=\chi_s+\chi_t+\chi_u$.}. The low-temperature phase is
symmetric in the moduli $s$, $t$ and $u$: it is common to the three
dual strings, in their perturbative and non-perturbative domains. In
contrast, the high-temperature heterotic phase only exists in the
perturbative domain of the heterotic string, where $s$ is the
dilaton, and, by duality, in non-perturbative type II regimes.

In the computation of the mass spectrum, one needs then to isolate the
contributions from the non-zero ${\cal G}_S$ in the mass matrices.
Because of the existence of couplings
$SUZ_2^2$ and $STZ_3^2$ in the superpotential, there will be
mass splittings of the O'Raifeartaigh type in the sectors $Z_2$
and $Z_3$. It turns out that all supersymmetry breaking contributions
to
the mass matrices are due to these superpotential couplings.
We then conclude that
the spectrum is supersymmetric in the perturbative
heterotic and moduli sector ($T,U,Z_1$), and with O'Raifeartaigh
pattern in the non-perturbative sectors:
$$
\begin{array}{rrcl}
Z_2:&  m_{bosons}^2 &=& m_{fermions}^2 \pm 2su\, m_{3/2}^2,
\crbig
Z_3:&  m_{bosons}^2 &=& m_{fermions}^2 \pm 2st\, m_{3/2}^2.
\end{array}
$$
As already observed in Ref. \cite{AK}, a similar analysis 
applied to the perturbative heterotic string only
would have led to a supersymmetric spectrum.

In the special infinite heterotic temperature limit
discussed in Ref. \cite{ADK}, in which $\alpha^\prime_H\to 0$, all
massive states decouple and consequently one recovers $N=2$
unbroken (rigid) supersymmetry in the effective (topological) field
theory of the remaining massless hypermultiplets.

\section{The high-temperature heterotic phase transition}

As we already discussed, the
high-temperature phase of $N=4$ strings is described by a
non-critical string with central charge deficit $\delta {\hat c}=-4$,
provided the heterotic string is in the weakly-coupled 
regime with $g_H^2 < g_c^2 = {\sqrt2+1\over2\sqrt2}$.
One possible
description is in terms of the (5+1) super-Liouville theory
compactified (at least) on the temperature circle with radius fixed
at the fer\-mio\-nic point 
$R=\sqrt{\alpha^\prime_H /2}$. The perturbative
stability of this ground state is guaranteed when there is at least
$\rm {\cal N}_{sc}=2$ superconformal symmetry on the world-sheet,
implying at least $N=1$ supersymmetry in space-time. However, our
analysis of the previous section shows that the boson--fermion
degeneracy is lost at the non-per\-tur\-bative level, even though 
the ground state remains supersymmetric.

An explicit example with $\rm {\cal N}_{sc}=4$ superconformal was
given in Ref. \cite{K, AFK}. It is obtained when together with the
temperature circle there is an additional compactified coordinate on
$S^1$ with radius $R_6=\sqrt{\alpha^\prime_H /2}$. These two circles
are
equivalent to a compactification on $[SU(2)\times SU(2)]_k$ at the
limiting value of level $k=0$. Indeed, at $k=0$, only the 6
world-sheet fermionic $SU(2)\times SU(2)$ coordinates survive,
describing a ${\hat c}=2$ system instead of ${\hat c}=6$ of
$k\to\infty$, consistently with the decoupling of four
supercoordinates, $\delta {\hat c}=-4$. The central charge deficit is
compensated by the linear motion of the dilaton associated to the
Liouville field, $\phi=Q^\mu x_\mu$ with $Q^2=1/(2\alpha^\prime_H)$
so that $\delta {\hat c}_L=8\alpha^\prime_HQ^2=4$.

Using the
techniques developed in Refs. \cite{ABK, AFK}, one can derive the
one-loop (perturbative) partition function in terms of the left- and
right-moving degrees of freedom on the world-sheet \cite{ADK}:
\beq
\label{pf}
\begin{array}{rcl}
Z^{\rm Liouv}[SU(2)\times SU(2)]_{k=0} &=&
\displaystyle{
{\Im\tau^{-1}\over\eta^6~{\ov\eta}^{18}} 
{1\over 8}\sum_{\alpha,\beta,{\ov \alpha},{\ov\beta},h,g}
(-)^{\alpha+\beta+\alpha\beta}
\theta^2\left[^{\alpha}_{\beta} \right]
\theta\left[^{\alpha + h}_{\beta + g} \right]~ }
\crbig && \hspace{1.1cm}
\displaystyle{\times\,
\theta\left[^{\alpha - h}_{\beta - g} \right]
{\ov\theta}\left[^{{\ov\alpha} +h}_{{\ov\beta}+g} \right]
{\ov\theta}\left[^{{\ov\alpha} -h}_{{\ov\beta} -g} \right]
{\ov\theta}^{14}\left[^{{\ov\alpha}}_{{\ov\beta}}
\right]}.
\end{array}
\eeq
This partition function encodes a number of pro\-perties, which
deserve some comments:
\begin{itemize}
\item
The initial $N=4$ supersymmetry is reduced to $N=2$ (or $N_3=4$)
because of the $Z_2$ projection generated by $(h,g)$. This agrees
with our effective field theory analysis of the high-temperature
phase given previously. The (perturbative) bosonic
and fermionic mass fluctuations are degenerate due to the remaining
$N=2$ supersymmetry.
\item
The $h=0$ sector does not have any massless fluctuation due to the
linear dilaton background or to the coupling to temperature. The linear
dilaton background shifts the bosonic masses (squared) by $m^2_{3/2}$,
so that all bosons in this sector have masses larger than or equal to
$m_{3/2}$. This is again in agreement with our effective theory
analysis. Similarly, fermion masses are shifted by the same amount
because of the $S^1_R$ temperature modification.
\item
In the $h=1$, ``twisted", sector there are massless excitations as
expected from the (5+1) super-Liouville theory \cite{BG, KS, AFK}.
\item
The $5+1$ Liouville background can be regarded as a Euclidean
five-brane solution wrapped on $S^1\times S^1$ preserving one-half of
the space-time supersymmetries ($N=2$).
\item
The massless space-time fermions in the $h=1$ sector are
six-dimen\-sio\-nal spinors constructed with the left-moving
su\-perco\-or\-dinates
$\Psi_{\mu}$ and $\beta,\gamma$ superghosts.
They are also spinors under the $SO(4)_{\rm right}$
constructed using four right-moving fermions ${\ov \chi}_I$ which
parametrize the fifth and sixth compactified coordinates,
with $R= R_6=\sqrt{\alpha^\prime_H/2}$. They are also
vectors under the $SO(28)$
constructed with 28 right-moving fermions ${\ov \Psi}_A$ ($c_R=14$).
\item
Similarly, the massless space-time bosons are
$SO(4)_{\rm right}$ spinors and $SO(28)$
vectors. They are also spinors under the $SO(4)_{\rm left}$
constructed with the left-moving fermions ${\chi}_I$ for
the fifth and sixth coordinates compactified at the fermionic point.
Together with the massless fermions,
they form 28 $N=2$ hypermultiplets.
\end{itemize}
These 28 massless hypermultiplets are the only states that survive in
the zero-slope limit and their effective field theory is described by
a $N=2$ sigma-model on a hyper-K\"ahler manifold. This topological
theory arises in the infinite temperature limit of the $N=4$
strings after the heterotic Hagedorn phase transition.

Although the $5+1$ Liouville background is perturbatively stable due
to the ${\rm \cal N}_{sc}=4$ superconformal symmetry, its stability
is not ensured at the non-perturbative level when the heterotic
coupling is large:
\beq
g^2_H(x_{\mu})=e^{2(\phi_0 - Q^{\mu}x_{\mu}) }>
{\sqrt{2}+1 \over 2\sqrt{2}} \sim 0.8536.
\eeq
Indeed, the
high-temperature heterotic phase on\-ly exists if $g^2_H(x_{\mu})$ is
lower than a critical value separating the heterotic and Type II
high-tem\-pe\-ra\-tu\-re phases. Thus one expects a domain wall in
space-time, at $x_{\mu}^0=0$, separating these two phases:
$g^2_H(Q^{\mu} x^0_{\mu})\sim 0.8536$. This domain wall problem can
be avoided by replacing the $(5+1)$ super-Liouville background with a
more appropriate one with the same superconformal properties, $\rm
{\cal N}_{sc}$ $ = 4$, obeying however the additional perturbative
constraint $g^2_H(x_{\mu})<<1$ in the entire space-time.

Exact superstring solutions based on gauged WZW two-dimensional
models with $\rm {\cal N}_{sc}=4$ superconformal symmetries have been
studied in the literature \cite{ABS, KPR, K, AFK, KKL}. We now
consider the relevant candidates with $\delta \hat{c}=-4$.

The first one is the $5+1$ super-Liouville with $\delta{\hat c}=4$,
already examined above. It is based on the $2d$-current algebra:
\beq
\begin{array}{c}
U(1)_{\delta\hat c=4} \times U(1)^3 \times
U(1)_{R^2={\alpha^\prime_H /2}} 
\times U(1)_{R_6^2={\alpha^\prime_H /2}}
\crbig 
\equiv ~U(1)_{\delta\hat c=4} \times U(1)^3 \times SO(4)_{k=1}.
\end{array}
\eeq

Another class of candidate background consists of the non-compact
parafermionic spaces described by gauged WZW models:
\beq
\begin{array}{c}
\displaystyle{\left[{SL(2,R) \over U(1)_{V,A} }\right]_{k=4} \times
\left[{SL(2,R) \over U(1)_{V,A} }\right]_{k=4} \times
U(1)_{R^2={\alpha^\prime_H /2}} \times  U(1)_{R_6^2={\alpha^\prime_H
/2}}}
\crbig \displaystyle{
\equiv
\left[{SL(2,R) \over U(1)_{V,A} }\right]_{k=4} \times
\left[{SL(2,R) \over U(1)_{V,A} }\right]_{k=4} 
\times SO(4)_{k=1},}
\end{array}
\eeq
where indices $A$ and $B$ stand for the ``axial" and ``vector"
WZW $U(1)$ gaugings.

Then, many backgrounds can be obtained by marginal deformations of
the above, preserving at least $\rm{\cal N}_{sc}=2$, or also by
acting with S- or T-dualities on them.

As already explained, the appropriate background must verify
the weak-coupling constraint:
\beq
\label{lastequ}
g^2_H(x_{\mu})=e^{2\phi}<<\,\sim \,0.8536 \,,
\eeq
in order to avoid the domain-wall problem, and in order to trust the
perturbative validity of the heterotic string background. This
weak-coupling limitation is realized in the ``axial" parafermionic
space. In this background, $g^2_H(x_{\mu})$ is bounded in the entire
non-compact four-dimensional space, with coordinates
$\{z,z^*,w,w^*\}$, provided the initial value of
$g^2_0=g^2_H(x_{\mu}=0)$ is small.
\beq
{1\over g^2_H(x_{\mu})}=e^{-2\phi}={1\over g^2_0}~(1+zz*)~(1+ww*)~\ge
{}~{1\over g^2_0}.
\eeq
The metric of this background is everywhere regular:
\beq
ds^2={4dzdz^*\over 1+zz^*} +{4dwdw^*\over 1+ww^*}\,\,.
\eeq
The Ricci tensor is
\beq
R_{z\,z^*}={1\over (1+zz^*)^2}~, \quad R_{w\,w^*}={1\over
(1+ww^*)^2}\,\,.
\eeq
The scalar curvature
$$
R={1\over 4(1+zz^*)}+{1\over 4(1+ww^*)}
$$
vanishes for asymptotically large values of $|z|$ and $|w|$
(asymptotically flat space). This space has maximal curvature
when $|z|=|w|=0$. This solution has a behaviour similar to that
of the Liouville solution in the asymptotic regime
$|z|,~|w| \rightarrow \infty$. In this limit, the dilaton $\phi$
becomes linear when expressed in terms of the flat
coordinates $x_i$:
\beq
\phi=-\Re[\log z]-\Re[\log w] = -Q^1|x_1|-Q^2|x_2|,
\eeq
where
$$
\begin{array}{rclrcl}
x_1&=&-{\Re}[\log z], \quad&
x_2&=&-{\rm Re}[\log w], \crbig
x_3&=&{\rm Im}[\log z], &
x_4&=&{\rm Im}[\log w],
\end{array}
$$
and the line element is $ds^2=4(dx_i)^2$. The important point here is
that, for large values of $|x_1|$ and $|x_2|$, $\phi\ll0$, in
contrast to the Liouville background in which $\phi=Q^1x_1+Q^2x_2$,
the dilaton becomes positive and arbitrarily large in one half of the
space, violating the weak-coupling constraint (\ref{lastequ}).

We then conclude that the high-temperature phase is described by the
above parafermionic space, which is stable because of $N=2$
supersymmetry. Since it is perturbative everywhere, the perturbative
massive bosonic and fermionic fluctuations are always degenerate.
On the other hand, 
the non-perturbative ones are superheavy and decouple in the
limit of vanishing coupling.

The asymptotic solution of the parafermionic space suggests an
alternative super-Liouville solution with
\beq
\begin{array}{rcl}
\phi &=& \phi_0-\Re[\log z]-\Re[\log w] \crbig
&=& \phi_0-Q^1|x_1|-Q^2|x_2|.
\end{array}
\eeq
The appearance of the absolute value of $|x_i|$ gives an upper bound
on the coupling constant provided $Q^i$ are positive. However, the
conical
singularity at $x_i=0$ implies, via the dynamical equation (\ref{dileom}),
the presence of curvature singularities at these points,
\beq
R_{z\,z^*}=-\pi\delta^{(2)}(z), \quad R_{w\,w^*}=-\pi\delta^{(2)}(w).
\eeq
In the above modified Liouville background, the $g^2_H(x_{\mu})$
is bounded in the entire
non-compact four-dimensional space,  provided the initial value
$g^2_0=g^2_H(x_{\mu}=0)=e^{2\phi_0}$ is small.

\section{Conclusions}

$N=4$ superstring theories at finite temperature $T$ correspond to
a particular gauging of the $N=4$ supergravity. Using techniques of
$N=4$ gauged supergravity, we were able to compute the exact effective
potential of all potential tachyonic modes, describing all three
perturbative instabilities of $N=4$ strings (heterotic, type IIA
and type IIB) simultaneously. Hagedorn instabilities of different
perturbative string descriptions appear as thermal dyonic 1/2-BPS
modes that become massless (and then tachyonic) at (above) the
corresponding Hagedorn temperature.

We find that the $N=4$ thermal potential has a global stable
minimum in a region where the heterotic string is weakly-coupled, so
that the four-dimensional 
string coupling $g^2_H< {\sqrt{2}+1 \over 2\sqrt{2}}$.
At the minimum, the temperature is fixed in terms of the heterotic
string
tension, the four internal supercoordinates decouple, and the system is
described by a non-critical superstring in six dimensions.
Supersymmetry,
although restored in perturbation theory, appears to be broken at the
non-perturba\-tive level.

On the heterotic or type IIA side, the high-temperature limit
corresponds to a topological theory described by an $N=2$
supersymmetric sigma-model on a non-trivial hyper-K\"ahler manifold.

\section*{\bf Acknowledgements}

This research has been partially
supported by the EEC under the contracts TMR--ERBFMRX--CT96--0045 and
TMR--ERBFMRX--CT96--0090, by the Swiss National Science Foundation and the
Swiss Office for Education and Science.

\end{document}